# Interfacial superconductivity in Cu/Cu$_2$O and its effect on shielding ambient electric fields[†]

Dale R. Harshman [a,*] and Anthony T. Fiory [b]

[a] *Physikon Research Corporation, Lynden, WA 98264-9335, USA*
[b] *Bell Labs Retired, Summit, NJ 07901-3437, USA*

**ABSTRACT**

A model is presented for two-dimensional superconductivity at semiconductor-on-metal interfaces mediated by Coulomb interactions between electronically-active interface charges in the semiconductor and screening charges in the metal. The junction considered is native Cu$_2$O on Cu in which an interfacial double charge layer of areal density $n$, comprising superconducting holes in Cu$_2$O and mediating electrons in Cu, is induced in proportion to a sub-monolayer of adsorbed $^4$He atoms. Evidence for superconductivity on copper with prior air exposure is revealed in new analysis of previously published work function data. Based on a theory developed for layered superconductors, the intrinsic transition temperature $T_C = \beta\, n^{1/2}/\zeta$ is determined by $n$ and transverse distance $\zeta \simeq 2.0$ Å between the charge layers; $\beta = 1.933(6)\, e^2 \lambdabar_C/k_B = 1247.4(3.7)$ K-Å$^2$ is a universal constant involving the reduced Compton wavelength of the electron $\lambdabar_C$. This model is applied to understanding the shielding of copper work-function patch and gravitational compression electric fields reported in the Witteborn-Fairbank gravitational electron free fall experiment. Interfacial superconductivity with $n \simeq 1.6 \times 10^{12}$ cm$^{-2}$, $T_C \simeq 7.9$ K and Berezinskiĭ-Kosterlitz-Thouless temperature $T_{BKT} \simeq 4.4$ K accounts for the shielding observed at temperature $T \simeq 4.2$ K. Helium desorption and concomitant decreases in $n$ and $T_C$ replicate the temperature transition in ambient electric fields on falling electrons, as observed by Lockhart et al., and the vanishing of superconductivity above $T \simeq 4.8$ K.

**Keywords:** 2D superconductor, Electron-mediation, Electron free fall, Metal-semiconductor interfaces, Helium adsorption.

---

[*] Corresponding author.
*E-mail address:* drh@physikon.net (D.R. Harshman).

---

## 1. Introduction

Two-dimensional superconductivity is hosted in thin layers and at interfaces of differing materials, being notably bounded by crossovers to bosonic pairing at reduced carrier density and elevated sheet resistance [1-11]. Recalling previously-discussed possibilities for a condensed Bose gas in an electron surface layer [12], this work presents a model for two-dimensional superconductivity in a semiconductor interfaced with a metal and mediated by Coulomb interactions with screening charges. The pairing mechanism is electronic, yet differing from earlier concepts of superconducting surfaces [13,14] and interfaces [3,15-20], and is based on interlayer Coulomb-mediation initially introduced for high-$T_C$ superconductivity [21]. While such an interfacial superconductive state portends intriguing materials physics, experimental data are available for testing the particular case of metallic Cu coated with its native oxide, forming a Cu/Cu$_2$O interface on the surface [22-24]. Evidence for superconductivity on previously air-exposed copper is manifested in a jump in the thermally-modulated work function, observed to be similar in magnitude to superconducting niobium (see Section 3.1). Surface superconductivity provides a plausible and quantifiable explanation for the shielding of ambient electric fields arising from work-function patch effects [25] and differential gravitational compression (DMRT field [26-29]) inside the copper tube held at 4.2 K in the electron gravitational drift measurements of Witteborn and Fairbank [30,31]. Copper-coated electrode plates in magnetic field levitation apparatus operating at liquid helium temperatures have also conveyed no evidence of patch-effect electric fields in contrast to a titanium coating [32,33].

Adsorption of $^4$He atoms, presumed to be in trace quantities, appears to be essential for producing these unique-to-copper effects. This interpretation follows from the resemblance of the temperature dependence of the ambient electric field reported by Lockhart et al. [34] to that of the Cu work function in $^4$He gas [35], as noted in an early review [36]. Quantitative analysis of these effects is presented in the experimental Section 3. It is, therefore, proposed herein that the Cu/Cu$_2$O interface junction is charged by a sub-monolayer of adsorbed $^4$He atoms. Strong electron repulsion by $^4$He, effectively pushing electron charge towards the copper, induces a double charge layer at the interface, comprising positively charged holes in semiconducting Cu$_2$O and negatively charged screening charges in the Cu. These charges form in equal numbers as a consequence of electrical neutrality of the $^4$He ad-atoms.

A schematic drawing of the theoretical Cu/Cu$_2$O structure is given in Fig. 1, which includes the adsorbed $^4$He atoms. Valence band holes in Cu$_2$O form a 2D superconductor mediated by their Coulomb interactions with the screening charges in the Cu. Image-potential band-gap narrowing at the interface [16,18,19,37] may also contribute to stabilizing the 2D hole layer. As previously shown for thin-film and atomic layer structures [38,39], pairing arises from interlayer Coulomb interactions between electronic charges in adjacent superconducting and mediating layers, where optimization of the superconductive state is identified with a balance of charges between the two layers [21], taken as inherent for Cu/Cu$_2$O.

Witteborn and Fairbank measured time-of-flight distributions for electrons emitted upward along the axis of a vertical copper drift tube in vacuum for various values of applied electric field, introduced by passing a d.c. current along the tube [30,31]. The results give an ambient vertical electric field magnitude of $(0.02 \pm 0.09)$ $m_e g/e$, where $m_e$ is the electron rest mass and which is consistent with cancellation of gravity by the Schiff-Barnhill effect of gravitational sag in copper electrons [40]. Speculative explanations were offered for the evidently complete shielding of the much larger ($>10^3$ $m_e g/e$) patch and DMRT electric fields [41,42]. Notably, theoretical calculations of shielding by metallic surface layers fall 1 to 2 orders of magnitude short of the experimental observation [12,43], with the possible exception of a condensed Bose gas comprising paired charges [12]. The Cu/Cu$_2$O interface is more likely to provide the minimum materials disorder for enabling 2D superconductivity [44], when compared to the exposed surface considered by Bardeen [43], owing to hydrocarbons, bound hydroxyl, water, and possibly cupric products, e.g., CuO [22,23] and reflecting earlier concerns [12,45]. Moreover, superconductivity in Cu$_2$O benefits from high hole mobility ($>10^4$ cm$^2$V$^{-1}$s$^{-1}$) at low temperature [46].

A metallic layer in the superconducting state could provide the observed shielding for an electron transiting the copper tube at 4.2 K, provided that this is below the temperature of the Berezinskiĭ-Kosterlitz-Thouless (BKT) transition $T_{BKT}$ [47-50], a temperature generally falling below the intrinsic transition temperature $T_C$ defined by a vanishing energy gap $\Delta(T_C) = 0$. Theoretical interfacial superconductivity is introduced for Cu/Cu$_2$O by the interaction between holes in the Cu$_2$O and Cu electrons with $T_C = \beta/\ell\zeta$ from earlier work [21]. Here $n \equiv \ell^{-2}$ is the area density of the superconducting charges, $\zeta$ shown in Fig. 1 is the transverse distance between the superconducting and screening charges, and $\beta = 1.933(6)\ e^2\lambdabar_C/k_B = 1247.4(3.7)$ K-Å$^2$ is a universal constant, where $\lambdabar_C = \hbar/m_e c$ is the reduced Compton wavelength of the electron and $k_B$ is the Boltzmann constant.

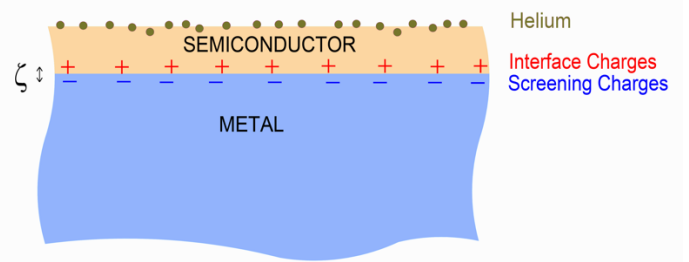

**Fig. 1.** Cross sectional view of the superconductor model structure of a semiconductor layer on a metal with adsorbed $^4$He atoms on the free surface, charges (+) in the semiconductor at the interface, screening charges (−) in the metal at the interface, and transverse separation distance $\zeta$ between the two types of charges.

The density $n$ of mobile holes, hence superconducting carriers, is assumed to be a fraction $f_s$ of the density of adsorbed $^4$He atoms, which obeys the thermally activated log-pressure law [51] and rapidly desorbs as temperatures are raised above 4.2 K. Modeling the temperature-dependent ambient electric field data in Section 3 yields $f_s \sim 0.06$. Parameter values at $T = 4.2$ K are a $^4$He fractional monolayer $\theta \simeq 0.034$, $n \simeq 1.6 \times 10^{12}$ cm$^{-2}$, $T_C \simeq 7.9$ K, and $T_{BKT} \simeq 4.4$ K. Magnitudes of these parameters decrease at elevated temperatures, leading to the vanishing of superconductivity above $T \simeq 4.8$ K. A non-negligible sub-monolayer of adsorbed $^4$He in the original work [30,31] may be surmised from the rather modest 80 L/s capacity of the ion pump, the protocol of not using the pump during data acquisition [41], and observations of $^4$He adsorption on copper at 4.2 K at pressures as low as $2 \times 10^{-11}$ Torr [52]. Although the temperature-dependence data were taken with a $^4$He gas leak (see Ref. [34:Erratum]), examination of $^4$He-electron scattering in Section 3.3 finds that the published ambient field data have plausible magnitudes. As discussed in Section 2.2 with application to data analysis in Section 3.4, the interfacial superconductor yields negligible shielding of electric fields from applied currents in the copper tube as well as external magnetic fields. These attributes are consistent with the null or at best inconclusive tests for superconductivity conducted with conventional macroscopic probes [45,53].

In Section 2, properties of the electronic structure are developed, coupled with Coulomb mediation, 2D superconductivity, and electrostatic screening effects. Section 3 presents analyses of experiments concerning evidence for superconductivity on copper, adsorption of $^4$He gas on copper, and temperature-dependent ambient electric fields observed for free-fall electrons. Section 4 offers discussions of key issues, followed by the Conclusion in Section 5.

## 2. Theory

The concept of superconductivity arising from interactions between charges in an insulator or semiconductor and screening charges in an adjacent metal is unique, and has the potential for augmenting the scope of interfacial superconductivity [3,11,44]. It follows as a direct extention of earlier theoretical work in which the genesis of unconventional superconductivity in certain layered compounds is non-phononic and vested in Coulomb interations between charges occupying adjacent charge reservoirs [21]. The shielding of electric fields by such a thin superconducting charge layer, while certainly of general interest, can also provide a quantitatively plausible explanation for the enhanced electric field shielding effects observed in early electron gravitational drift measurements [30,31,34]. In Section 2.1, a model is constructed based on the theory described in Ref. [21] and the electric field screening effects are evaluated in Section 2.2.

### 2.1. Superconductive model

Interfacial band gap narrowing arising from screened image potentials [37,54] appears to be applicable for the Cu/Cu$_2$O interfaces, based on analysis of Schottky diodes that position the Fermi energy at almost midway in the Cu$_2$O band gap [55]. Image-potential raising of the valence band edge at the interface may be a factor in forming an interfacial potential well for holes in thin-film Cu-Cu$_2$O structures. A further consideration is whether band closure or metal-induced gap states (MIGS) occur at the interface [18,19,56], such as been observed for Cu/MgO [57], even though the 2.1-eV band gap of Cu$_2$O is comparatively smaller. A MIGS-based interfacial superconductor has also been proposed, based on the idea that pairing can be mediated by excitons in the semiconductor [19].

Several experimental studies searching for evidence of superconductivity [45,53] have recognized that shielding by a two-dimensional superconductor is limited by the critical temperature $T_{BKT}$ [47-50], which from experiment [30] must be at least 4.2 K. Assuming disorder is a small perturbation, $T_{BKT}$ scales with $n/m^*$ for areal carrier density $n$ and effective mass $m^*$ and thus also with the Fermi energy $E_F = \pi\hbar^2 n/m^*$ as $T_{BKT} \sim 0.1\ E_F/k_B$, approaching the theoretical limitation [4,8]. An order of magnitude estimate for $T_{BKT} \sim 4.2$ K is $n/m^* \sim 10^{12}$ cm$^{-2}$ $m_e^{-1}$. Examination of the magnitude of $n/m^*$ for a given model can therefore reveal whether superconductivity is thermodynamically stable. In their idealized electron layer on a copper surface, Hanni and Madey [12] found that shielding for $n = (4$ to $5) \times 10^{10}$ cm$^{-2}$ (with $m^* = m_e$) hypothetically supports superconductivity at $T_{BKT} \approx 0.1$ K. The surface-state band on Cu$_2$O considered by Bardeen with hole concentration $n \approx 5 \times 10^{12}$ cm$^{-2}$ has a density-of-states $m^* \approx 170\ m_e$ [43], yielding $n/m^* \approx 3 \times 10^{10}$ cm$^{-2}$ $m_e^{-1}$ and $T_{BKT} \approx 0.1$ K. Although the values of $n/m^*$ derived in these quite different models for charge states on the surface appear to be too small by a factor of about 40, seemingly negating the possibility of superconductivity, the magnitudes were designed to provide screening that replicates the ambient electric fields observed at temperatures above 4.2 K [34].

Adsorption of $^4$He on the native oxide surface of Cu evidently induces $n/m^*$ sufficient for superconductivity at 4.2 K, which is deduced in the following from the change in work function $\Delta W = -6$ meV measured at 4.2 K for an approximately monolayer coverage [35]. The sign of $\Delta W$ is opposite to the approximately $+4$ eV measured for $^4$He adsorption on clean Cu [58], where electric dipoles oriented away from the surface develop at the adsorption sites [59-61]. Owing to strong electron repulsion, $^4$He effectively pushes negative charge towards the metal [59,60], which is presumed to be the metallic side of the Cu/Cu$_2$O interface in the presence of native oxide. High Allen-scale $^4$He electronegativity of 4.160, when compared to O (3.610), N (3.066), C (2.544), and Cu (1.85), expresses quantitatively this unique distinction of the full-shell valence electrons in $^4$He. The double charge layers depicted in Fig. 1 follow from interpreting the observed negative $\Delta W$ as $^4$He pushing some charge across the interface, promoting the formation of screening electrons in the Cu and a like number of hole carriers in the Cu$_2$O. The fractional charge per $^4$He atom is found to be small ($f_S \approx 0.06$, Section 3.4) and thus entails small perturbations of pre-existing charges at the Cu-Cu$_2$O interface. In a rudimentary approach considering $-\Delta W$ as the shift in potential produced by filling a 2D electronic hole band, the free areal charge density $n = -\Delta W D$ for density of states $D = m^*/\pi\hbar^2$ yields $n/m^* = -\Delta W/\pi\hbar^2 = 2.75 \times 10^{12}$ cm$^{-2}$ $m_e^{-1}$. This number is actually the lower bound on $n/m^*$, given that contributions to $\Delta W$ from outwardly oriented electric dipoles at the adsorption sites are unknown. However, the value shows compliance with the aforestated criteria for minimum $n/m^*$, confirming that superconductivity with $T_{BKT} \gtrsim 4.2$ K is plausible. One may note that the magnitude of $\Delta W$ is on the order of 0.1% of $W \approx 4.65$ eV for clean polycrystalline Cu [62]) and ~1% of the $+0.5$ eV increase upon exposure of clean Cu to a dry O$_2$-N$_2$ ambient [63]; an increase with oxidation is ascribed to the electronegativity of O [64].

Accepting that Bardeen's analysis yields the correct order of magnitude for $n$ [43], the foregoing result predicts that $m^*$ is on the order of $m_e$ and thus comparable to a Cu$_2$O-band effective mass. A plausible value $m^* = 0.69 m_e$ is that measured for upper valence band holes (Cu 3$d$ levels) at low temperatures, which exhibit high mobility ($5 \times 10^4$ cm$^2$V$^{-1}$s$^{-1}$) in bulk crystals of Cu$_2$O [46]. For the transverse distance $\zeta$ in Fig. 1, the charges are taken to occupy two planes defined by the locations of interfacial Cu ions. Averaging the ionic positions reported in several theoretical structure calculations of Cu/Cu$_2$O interfaces [24,65,66] gives $\zeta = 2.0$ Å.

Simulations of oxidized Cu using an atomically sharp Cu(100)/Cu$_2$O(111) interface [24] find positive and negative relative charges up to about $\pm 0.2 |e|$ on Cu in one to two layers into the metal, indicating built-in charge densities of $\pm 10^{13}/|e|$ cm$^{-2}$ that are slightly net negative. Magnitudes of these localized charges are comparable to the surface charge on a metal induced near a test charge at distance $\zeta/2$, according to the idealized method of images. Estimating characteristic lengths of such a surface charge in linear response theory [67],[1] the centroid is located 0.7 Å closer to the interface, relative to the edge of a uniform positive-charge background, and is of thickness $\approx 2.2$ Å. The negative charges idealized in Fig. 1 therefore have their genesis in the net negative charges on interfacial Cu atoms.

Superconductivity at surfaces and interfaces, mediated solely by electronic or excitonic coupling, has been explored theoretically for some time in various contexts [15-18,20,68]. Electronic superconductive pairing in a semiconductor adjoining a metal, based on Coulomb interactions with screening electrons formed in the metal [12], appears possible because the charge layers shown in Fig. 1 directly reflect the elementary structural unit of a high-$T_C$ superconductor [21,39]. The theoretical intrinsic transition temperature is defined by,

$$T_C = \beta/\ell\zeta = k_B^{-1}(\Lambda/\ell)\,(e^2/\zeta)\,, \qquad (1)$$

where $\ell$ is determined by the areal density of superconducting charges $n \equiv \ell^{-2}$ and $\zeta$ is the perpendicular distance between the charge layers ($\beta = 1247.4(3.7)$ K-Å$^2$ and $\Lambda = 1.933(6)\,\lambda_C$). According to [21], optimization of the superconductive state requires balance between the two types of charges, which is intrinsically satisfied in this case.

For a two-dimensional superconductor, the topological phase transition temperature is given by $T_{BKT} = [c^2\hbar^2/16e^2k_B]/L_s = [1.961$ cm-K$]/L_s$, where $L_s$ is the renormalized magnetic screening length at temperature $T_{BKT}$ [50], and where $T_{BKT} < T_C$. Ambient electric field data [30,34] indicate $T_{BKT} \gtrsim 4.2$ K at $T \simeq 4.2$ K, for which $L_s \lesssim 0.47$ cm.

The ratio of $L_s$ to the intrinsic sheet magnetic penetration depth $\Lambda_0(T_{BKT})$ defines the renormalization factor $\varepsilon_C = L_s/\Lambda_0(T_{BKT})$, which is non-universal. The result $\varepsilon_C = 1.2 \pm 0.1$ is obtained for superconducting films with high sheet resistance in Ref. [69] and is adopted here. A superconducting layer sheathing a copper tube of surface area 0.15 m$^2$ would comprise a mosaic of about 10$^4$ regions of area $L_s^2$ that could sustain long range phase coherence at zero applied current. Consequently, the possibility exists for such a sheath to shield the patch and gravitational stress electric fields generated internally in the copper, while also freely admitting externally applied electric fields.

---

[1] As described by Bardeen, "These charges arise from a modification by the external field of the tail of the wave functions of the electrons extending into the vacuum" [43].

Temperature dependence in the intrinsic (*i.e.* not renormalized) penetration depth, expressed functionally as $f(T/T_C) \equiv \Lambda_0(0)/\Lambda_0(T)$ [70], is accurately described by the two-fluid expression $f(T/T_C) = 1 - (T/T_C)^4$ applicable to strongly coupled superconductors [71]. The renormalized screening length at $T = T_{BKT}$ is thus expressed as $L_s = \varepsilon_C \Lambda_0(0)/f_C$, where $f_C \equiv f(T_{BKT}/T_C)$. Effects of disorder, considered as possible in Ref. [42], are taken into account by using the Ginzburg Landau form $\Lambda_0(0) = \Lambda_L(1+ 0.75\,\xi_0/l)$ [70], where $\xi_0$ is the Pippard coherence distance, $l$ is the transport mean free path, and $\Lambda_L = m^*c^2/2\pi n e^2$ is the London sheet magnetic penetration depth for carriers of effective mass $m^*$ and areal density $n$. Circular energy contours are assumed, giving the expression,

$$T_{BKT} = \pi f_C\, n\hbar^2\, (8\varepsilon_C m^* k_B)^{-1} (1 + 0.75\,\xi_0/l)^{-1}. \qquad (2)$$

For a clean superconductor with $l \gg \xi_0$, Eq. (2) predicts the scaling $T_{BKT} \propto n/m^*$ with a non-universal numerical factor. A prediction from Eq. (2) is that the ratio of $T_{BKT}$ to the Fermi temperature $T_F = E_F/k_B$ is given by the expression,

$$T_{BKT}/T_F = f_C\,[8\varepsilon_C(1 + 0.75\,\xi_0/l)]^{-1}, \qquad (3)$$

which approaches the theoretical maximum $T_{BKT}/T_F \to 1/8\varepsilon_C$ in the limits $\xi_0/l \to 0$ and $f_C \to 1$. This upper bound originates from the universal jump in the renormalized superfluid density $\rho_S$ at $T_{BKT}$ [4,72].

The right side of Eq. (2) may be expressed in terms of the normal electrical sheet resistance $R_N = m^* v_F/ne^2 l$, the zero-temperature energy gap $\Delta_0 = \hbar v_F/\pi \xi_0$, where $v_F$ is the Fermi velocity, and the superconducting quantum resistance for pairs $R_Q = h/4e^2$ such that,

$$T_{BKT} = \pi f_C\,(4\varepsilon_C k_B)^{-1}(l/\xi_0 + 0.75)^{-1}(R_Q/R_N)\,\Delta_0. \qquad (4)$$

It is recognized that the penetration depth in the "dirty limit" $l \ll \xi_0$ is appropriate for very thin films of metallic superconductors, owing to high $R_N$ [73], giving $T_{BKT} \approx \pi f_C\,(4\varepsilon_C k_B)^{-1}(R_Q/R_N)\,\Delta_0$ in a form applicable to granular superconducting films [69]. Resistance $R_Q$ roughly sets the demarcation in $R_N$ between the superconducting and insulating phases [44,74-76]. Granular films comprise discontinuous superconducting grains interconnected by Josephson tunneling through weak links [76] coupled by Josephson critical current $I_C(T) \propto \Delta(T)/R_N$ [77,78], where $I_C(T) \propto (T_C - T)^2$ for insulating links [79], giving $k_B T_{BKT} = I_C(T_{BKT})h/8\varepsilon_C e$ [78], assuming negligible corrections for capacitive charging [80]. Although $R_N$ near $R_Q$ is considered in the present case, the superconducting layer is treated as a continuous two-dimensional sheet, rather than granular, because native oxidation produces an interface that completely covers the original Cu surface.

The transition temperature enters in Eqs. (2) – (4) through the expression $\Delta_0 = \gamma k_B T_C$ for coupling parameter $\gamma$. A value for low carrier density may be estimated from the quasi-2D superconductor Li$_{0.011}$ZnNCl [6] for which the tunneling gap $\Delta_0 = 4.50$ meV and intrinsic resistive $T_C = 20.1$ K work out to indicate strong coupling with $\gamma = 2.5$. Adopting this value here, the Pippard coherence distance $\xi_0 = \hbar v_F/\pi \gamma k_B T_C$ with Eq. (1) for $T_C$ gives $\xi_0 = \gamma^{-1}(2/\pi)^{1/2} \hbar^2 \zeta/\beta k_B m^* = 65.6$ Å. Note that this expression for $\xi_0$ is independent of $n$, $T_C$, and $R_N$.

Denoting the reduced normal-state resistance as $r_N = R_N/R_Q$ and given by $r_N = 4(2\pi)^{-1/2}(\ell/l)$, the mean free path may be expressed as,

$$l = 4\,(2\pi)^{-1/2}\,\ell/r_N, \qquad (5)$$

showing that the limitation $r_N \lesssim 1$ for superconductivity corresponds to $l \gtrsim 1.596\,\ell$. The criterion $k_F l \gtrsim 1$ for a metallic normal state of Fermi wavevector $k_F = (2\pi)^{1/2}/\ell$ evaluates to $k_F l \gtrsim 4/r_N$, which is automatically satisfied for $r_N \lesssim 1$. Equation (5) is utilized by assuming that $r_N$ is a materials constant independent of the quantity of adsorbed $^4$He, which is rationalized by considering $^4$He as both effectuating $n$ as well as being dominant in carrier scattering. The screening charges function similarly to ultra-shallow ionized acceptors, yielding $l$ scaling with $\ell$ at low temperature.

### 2.2. Effect of electrostatic fields

The following treats the function of the electron beam circuit as a current detector connected to the copper tube and thereby the superconducting surface layer. This emulates the electrical system of the apparatus [41], where the copper tube is resistively connected to the electron source and detector through external power supplies. Considering the case of zero current in the copper tube, the electrochemical potential in the superconductor is given to be nearly uniform and constant in time, supporting counter-flowing supercurrents that cancel electrostatic fields [81,82], including the DMRT and patch (surface-parallel component) electrostatic fields introduced by the copper tube. Temperature gradients along the tube are presumed to be small enough for cancelling thermoelectric fields as well. As a consequence of the supercurrent being confined to a curved two-dimensional surface, components of electrostatic fields perpendicular to the tube surface remain unshielded.

Superconductivity is affected by small surface-normal components of the longitudinal magnetic field used for centering the electron flow along the tube axis [41]. Such locally transverse magnetic field components $B_\perp$, which arise from variations in magnitude and axial alignment [41], generate free supercurrent vortices (magnetic flux quanta) with densities and both senses of circulation changing sinusoidally around the circumference of the tube. Guided by planned magnetic field variations up to 0.1 G [41] and for $B_\perp$ assumed to be within ±0.05 G, the average free vortex density is $n_f = (1/2)B_\perp/\varphi_0 \approx 10^5 \text{cm}^{-2}$, where $\varphi_0$ is the superconducting flux quantum, corresponding to $N_v \approx 10^8$ vortices in the superconductor (a cylinder of diameter $d_T = 5$ cm and length $h = 90$ cm [41]). Free vortices in a superconductor are generally held stationary by pinning at defects (e.g., related to the copper surface texture [41]), thereby keeping the superconducting phase invarient with time. Temperature gradients along the tube are presumed to be sufficiently small so as to maintain zero magneto-thermoelectric voltages in the pinned vortex state [83,84]. Thus, the screening effect associated with the superconductivity is time invariant.

Exceptions to pinned vortices occur during flux flow driven by a transport current at temperatures very close to $T_{BKT}$, where thermally excited vortex-pairs screen the pinning centers, allowing the free vortices to produce flux flow resistance and dissipation [85]. From the Bardeen and Stephen theory of vortex motion [86], the sheet resistance is given by $R_f = 2\pi\xi^2 B_\perp R_N/\varphi_0$, which evaluates to 0.014 Ω by taking $\xi = 0.855 [\xi_0 l/(1 - T_{BKT}/T_C)]^{1/2} = 135$ Å from the "dirty" form [70]. The resistance of the superconducting cylinder may be approximated as $R = R_f h/\pi d_T = 0.08$ Ω, which generates a Johnson-Nyquist noise voltage $V_{rms} = (4k_B TR/\Delta t)^{1/2}$. With $\Delta t \sim 1$ second as the transit time for the slowest electrons [41], one has $V_{rms} = 4\times 10^{-12}$ V.

Microscopically, thermal noise originates from random vortex motion with diffusivity $D_v = \mu_v k_B T$ [50], as expressed terms of the vortex mobility $\mu_v = (R_N/R_O)\xi^2/\hbar$ derived from $R_f$. At $T = 4.2$ K, $D_v = 0.80$ cm$^2$/s. Diffusive displacement of a single vortex projected along the circumference of the cylinder changes the phase of the superconducting order parameter along its length by $\Delta\phi \sim (2D_v\Delta t)^{1/2}/\pi d_T \approx 0.5$ radian and, from the a.c. Josephson equation, a voltage noise increment of $V_v = (\hbar/2e)\Delta\phi/\Delta t \approx 2 \times 10^{-16}$ V. Average voltage noise created by $N_v$ vortices is thus estimated from the standard deviation as $\pm N_v^{1/2} V_v \sim \pm 2 \times 10^{-12}$ V, in essential agreement with $V_{rms}$. Vortex pinning emerges for $T < T_{BKT}$ as the number of thermally excited vortex pairs rapidly diminishes, decreasing the flux-flow resistance [85] and accordingly the thermal voltage noise. The above estimates are comparable to the total experimental noise, which is at least 0.09 $mgh/e \approx 5\times 10^{-12}$ V, expressed as a voltage [30].

An applied electric field $E_A$ is induced by passing an electric current longitudinally through the copper tube. According to specifications in Ref. [41], for example, $E_A = 10^{-10}$ V/m is obtained with a current of $2 \times 10^{-4}$ A through a tube of resistance $4.5 \times 10^{-7}$ Ω. Continuity of $E_A$ across the copper-superconductor interface is maintained by the flux-flow resistive state. For $R \sim 0.08$ Ω, the superconductor transports a fraction of at least $6 \times 10^{-6}$ of the applied current, exclusive of an indeterminate vortex de-pinning current. The DMRT and patch fields are thermodynamically non-dissipative, e.g., the lattice compression field in the copper is largely shielded (factor ~1/7 [27]) by mobile electrons in the copper. Hence, these electrostatic fields neither invoke nor contribute to time dependence in the phase of the superconducting order parameter, such as flux-flow dissipation, in the presence of a transport current.

The ambient electric field $E$ inside the copper tube (positive denoting the upward direction) is the sum of components:

$$E = E_G + SE_T + E_A, \qquad (6)$$

where $E_G = -m_e g/e = -5.6 \times 10^{-11}$ V/m is the gravitational sag field from mobile carriers [40], $E_T$ is the sum of the gravitational and patch fields from the copper tube, $S$ is the shielding factor, and $E_A$ is the electric field applied to the copper tube. For mobile carriers in a normal 2D Fermi gas, the shielding factor is $S_N = (\partial\mu/\partial n)/4\pi e^2 \zeta$ with chemical potential $\mu = k_B T \ln[\exp(\pi\hbar^2 n/m^* k_B T) - 1]$ [12]. Temperature dependence in $S_N$ is given by,

$$S_N = (\hbar^2/4m^* e^2 \zeta) [1 - \exp(-\pi\hbar^2 n/m^* k_B T)], \qquad (7)$$

where the coefficient sets the low-temperature limit as $S_N \rightarrow \hbar^2/4m^* e^2 \zeta = 0.096$. Normal excitations in the superconducting state are assumed to have the same $S_N$ of Eq. (7) as in the normal state.

Superconductivity is introduced by using the two-fluid model with normal and superconducting shielding factors, denoted $\tilde{S}_N$ and $\tilde{S}_S$, respectively. Drawing an analogy to parallel transport in the two-fluid model, superconductivity in the surface electron layer is constructed as a parallel connection of two resistances, $\tilde{R}_N$ and $\tilde{R}_S$, representing the normal and superconducting components, respectively, which are coupled to the copper tube by a series resistance $R_C$. For voltage drop $V_T$ across the length of the copper tube and voltage drop $V_{OUT}$ across the length of the surface electron layer, the ratio of the voltages gives the shielding factor as $S = V_{OUT}/V_T = [1 + R_C(\tilde{R}_N^{-1} + \tilde{R}_S^{-1})]^{-1}$. The normal shielding factor is obtained from $S$ in the limit $R_S \to \infty$ as $\tilde{S}_N = (1 + R_C/\tilde{R}_N)^{-1}$ and the superconducting shielding factor in the limit $R_S \to 0$ as $\tilde{S}_S = \tilde{R}_S/R_C$. The shielding factor is thus obtained in terms of normal and superconducting components by the expression,

$$S = (\tilde{S}_N^{-1} + \tilde{S}_S^{-1})^{-1} = \tilde{S}_N/(1 + \tilde{S}_N/\tilde{S}_S) . \qquad (8)$$

In the normal state $T > T_C$, $S = S_N$ and $\tilde{S}_S = 1$, thereby determining $\tilde{S}_N = S_N/(1 - S_N)$ in terms of $S_N$ in Eq. (7). In the fully coherent superconducting state, $\tilde{S}_S/\tilde{S}_N \ll 1$, yielding $S = \tilde{S}_S$. Minimum values of $\tilde{S}_S$ in the superconducting state are possibly limited by finite critical currents and extrinsic disorder.

The ambient electric field in the presence of superconducting shielding thus evaluates to,

$$E = E_G + \tilde{S}_N E_T /(1 + \tilde{S}_N/\tilde{S}_S) + E_A . \qquad (9)$$

The electronic gravitational sag field term $E_G$ acts against gravity, irrespective of the presence of superconductivity [40]. The substances coating the copper tube are considered to have insufficient mass or mechanical rigidity for generating the theoretical DMRT field of a bulk insulator [28].

At temperatures $T > T_{BKT}$, the presence of superconducting fluctuations is assumed to yield partial shielding, such that in Eq. (9) one has $\tilde{S}_S^{-1} \propto \xi_+^2$, where $\xi_+$ is the order parameter correlation length. This assumes that $\tilde{S}_N^{-1}$ scales in like manner as the fluctuation conductivity and transverse magnetic susceptibility, both of which vary as $\xi_+^2$. For $T \gtrsim T_{BKT}$, $\xi_+^2 \propto \exp\{2[b\tau_C/(T/T_{BKT} - 1)]^{1/2}\}$, where $b$ is a dimensionless parameter of order unity and $\tau_C = T_C/T_{BKT} - 1$ [50]. The maximum value taken by $\xi_+$ is limited to the lesser of $L_s$ or the spacing between free vortices spawned by an external magnetic field. Using the interpolation formula for $T > T_{BKT}$ presented in Ref. [50], fluctuation shielding is approximated for temperatures up to $T_C$ by,

$$\tilde{S}_S = S_1 b \sinh^{-2}\{[b\tau_C/(T/T_{BKT} - 1)]^{1/2}\} \qquad (10)$$
$$(\text{for } T_{BKT} < T < T_C) .$$

For $T \gtrsim T_C$, solution of Eq. (10) for $\tilde{S}_S = 1$ establishes the value of the prefactor $S_1 = b^{-1} \sinh^2\{[b\tau_C/(T_C/T_{BKT} - 1)]^{1/2}\}$.

## 3. Experiment

Analyses of published experimental data for oxidized copper are presented in the following sub-sections, showing the evidence for superconductivity, a 2D electronic layer, and temperature-dependent superconductive shielding of ambient electric fields, while also quantitatively relating the experimental observations to $^4$He adsorption.

### 3.1 Evidence for superconductivity

Work-function temperature-derivatives $dW/dT$ vs. $T$ for native-oxidized polycrystalline copper were reported by Free et al. [87]. Figure 2 shows data transcribed from Figs. 3 and 4 therein, which has extended temperature range compared to earlier works [35,88]; $W$ is expressed in the experimental units of voltage. The study was intended to search for features that could be associated with a shielding transition [34], yet concluded that no anomalies were observed, e.g., at 4.2 or 4.5 K. However, the temperature dependence of $dW/dT$ does indeed contain evidence of a discontinuity at the lower temperature of $T_X = 3.7$ K, as indicated by two solid red lines in Fig. 2 as described below. This feature is possibly related to the sharp change at approximately 4 K reported for the contact potential of copper in helium gas [89].

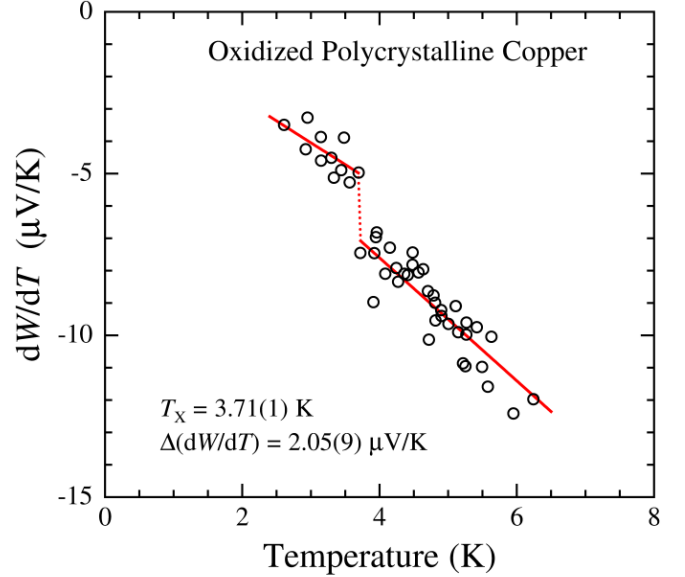

**Fig. 2**. Temperature derivative of the work function of copper with native oxide, $dW/dT$ (symbols), vs. temperature [87]. Lines denote linear-in-$T$ variations for $T < T_X$ and $T > T_X$ with jump $\Delta(dW/dT)$ indicated by dotted line at $T_X$. Values of parameters in legend.

The solid lines in Fig. 2 denote the function

$$dW/dT = w_1\theta_v(T_X - T)T + w_2\theta_v(T - T_X)T, \quad (11)$$

where $\theta_v$ is the unit step function; a dotted line marks the discontinuity. A three-parameter regression fit to the data, satisfying the thermodynamic limit of $dW/dT \to 0$ as $T \to 0$, yields slopes $w_1 = -1.35(6)$ μV K$^{-2}$ and $w_2 = -1.90(2)$ μV K$^{-2}$, $T_X = 3.71(1)$ K, an rms deviation with respect to the data of 0.616 μV K$^{-1}$, and chi-square $\chi^2 = 0.380$ (for $\pm 1$ μV K$^{-1}$ intrinsic noise [87]). The jump $\Delta(dW/dT)$ at $T_X$ is defined by the discontinuity as $(w_1 - w_2)T_X = 2.05(9)$ μV K$^{-1}$. Division by the value $w_2 T_X = -7.0(3)$ μV K$^{-1}$ for $dW/dT$ at $T$ just above $T_X$ yields the dimensionless normalized jump given by the quantity $w_1/w_2 - 1 = -0.29(3)$. Several continuously smooth functions through the origin are also tested for statistical comparison. Fitting a single continuous line more than doubles $\chi^2$ to 1.027, while a smooth quadratic curve increases $\chi^2$ to 0.629, from which F-test probabilities of 0.0006 and 0.048 [90] are, respectively, obtained. Existence of a jump in the data for $dW/dT$ is therefore at least 95% probable, with its value resolved to better than 5% statistical accuracy.

This may be compared to results from work function data for Nb, where a jump in $dW/dT$ of about $-5$ μV K$^{-1}$ for $T_C = 9.2$ K was reported [88]. With $dW/dT \approx 15$ μV K$^{-1}$ just above $T_C$, estimated from Fig. 5 in Ref. [88], the normalized jump for Nb is about $-0.33$. Values of the normalized jump in $dW/dT$ for Cu and Nb therefore have the same sign and nearly equal magnitudes within the experimental uncertainties. In the case of Nb, the jump at $T_C$ is associated with superconductivity in Nb [88].

It has been suggested that the work function may be able to detect temperature dependent changes in the chemical potential $\mu$ at low electron density [91], particularly the jump in $d\mu/dT$ at $T_C$ that is derived in three-dimensional models [91]. While the behavior of $\mu$ in 2D and 3D may be similar [1,92], the signature feature of the superconducting transition in 2D generally occurs in the renormalized superfluid density $\rho_S$. Theory predicts a square-root singularity and discontinuity in $\rho_S$ at $T_{KBT}$ [50,72], which is observed to be continuous in temperature at finite frequency [85]. How such properties are reflected in measurements of $dW/dT$ remains an open question.

An electronic surface layer on the oxidized Cu may be assumed [12], since the measured magnitudes of $dW/dT$ are 50–100 times larger than estimates for bulk Cu and ~30 times larger than the measured thermopower of the sample [87]. As a surface probe, the work function is specific to electronic charge and presumably insensitive to mass or spin, hence it is logical to identify $T_X$ as an actual measurement of $T_{BKT} = 3.70(2)$ K. With parameters $m^* = 0.69 m_e$, $\zeta = 2.0$ Å, $r_N = 1.0$, and $\varepsilon_C = 1.2$ from Section 2.1, one obtains 2D carrier density $n = 1.29(1) \times 10^{12}$ cm$^{-2}$ and $T_C = 7.08(2)$ K at $T_X$. Other derived parameters with values at $T_X$ are $k_F\xi_0 = 1.86(1)$ and Fermi temperature $T_F = 52$ K, the latter yielding $T_{BKT}/T_F = 0.071$. Although care was taken in Ref. [87] to avoid introducing helium, vacuum pressure is not specified ($P = 10^{-7}$ Torr is stated for the Nb experiment [88]). A small sub-monolayer of $^4$He is therefore consistent with an unavoidable level of contamination as originally conjectured for high vacuum [31],[2] estimated from data analyses in Sections 3.2 and 3.4 to be 0.025 fractional $^4$He monolayer at 3.7 K.

The jump in the temperature dependence of the work function derivative $dW/dT$ measured for copper [87] is therefore considered to be significant evidence for a 2D superconducting transition on Cu. Interpretation of the jump's magnitude awaits further theoretical development, in particular for the temperature dependence in the chemical potential $\mu(T)$ in 2D. A jump in $\mu(T)$ at $T_C$ is derived for 3D models of low electron density superconductors, e.g. a magnitude of $\Delta_0^2/4T_C\mu_N(T_C)$, where $\mu_N(T_C)$ is the normal-state chemical potential at $T_C$ [91].

Temperature linearity in $dW/dT$ for $T < T_X$ points to the presence of a non-superconducting metallic phase on the surface, possibly as a coexisting bosonic metal [5,7,10,11]. However, the metallic signature may simply reflect sensitivity to the Cu substrate and the screening charges. Other possibilities are the surface states on Cu$_2$O considered by Bardeen [43] and/or surface charges [12,42], which could account for the linear-like behavior noted for the temperature dependence of ambient electric fields for $T > 4.2$ K [12,43].

## 3.2. Adsorption of $^4$He on copper

De Waele et al. [35] report that exposure of a Cu sample to helium ($^4$He) gas reduces the work function $W$ and present results (in units of voltage) for the temperature dependence of the change relative to 4.2 K, denoted as $W_T - W_{4.2}$. A digitized replica of the curve in Fig. 9 therein is shown as the solid curve in Fig. 3. A He leak produces a similar effect on the work function of Nb, indicated by comparable temperature derivatives $dW/dT$ at 4.2 K, i.e. ~3 mV/K for Cu and 1.6 mV/K for Nb [35]. Absent the admitted helium gas, results for $dW/dT$ have substantially smaller magnitudes (~1%): $-(8$ to $10)$ μV/K for Cu [35,87] and $+(8$ to $13)$ μV/K for Nb [35,88]. The following analysis considers the published drawing as accurately representing the data.

---

[2] See Ref. [31] (p. 1050), "A possible mechanism for this is adsorption of hydrogen and helium from the background gas during cooling." The concentration of $^4$He in air is 5.2 ppm.

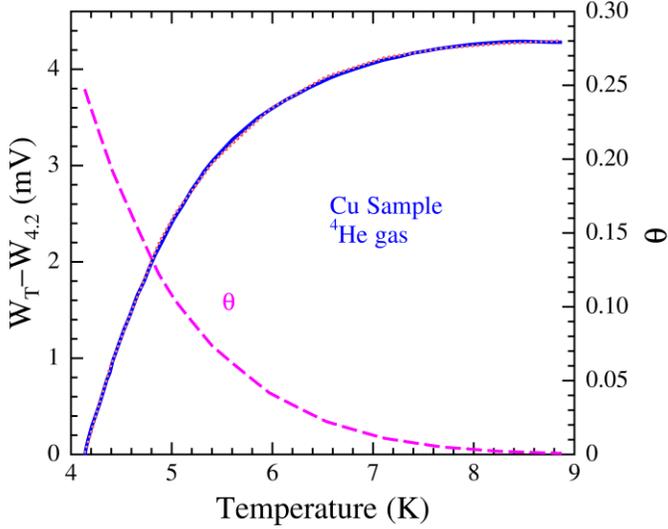

**Fig. 3.** Temperature dependence of the work function change $W_T - W_{4.2}$ for a copper sample in $^4$He gas (solid curve) [35] and model function (dotted curve). Curve $\theta$ (dashed, right scale) is the modeled fractional $^4$He gas coverage determined from Eq. (14).

The dotted curve in Fig. 3 is a function that scales $W_T - W_{4.2}$ with the fraction of $^4$He monolayer coverage $\theta$ [93] such that,

$$W_T - W_{4.2} = \Delta W + W_1 \theta, \quad (12)$$

where $\Delta W$ is the maximum work function change and $W_1$ (<0) is the rate of work function change per unit $\theta$. As reported in Ref. [35], the change in $W$ saturates at 6 mV for $\theta \sim 1$, which corresponds to a $^4$He areal density of about $7.9 \times 10^{14}$ cm$^{-2}$ [94]. Theory for $\theta$ is based on the log-pressure expression derived for gas adsorption on a flat surface [51],

$$\theta = \ln\{(U_b/k_B T)/\ln[P/P_{\text{vap}}(T)]\}, \quad (13)$$

where $U_b$ is the binding energy ($U_b < 0$), $P$ is the pressure, and $P_{\text{vap}}(T)$ is the saturated vapor pressure at temperature $T$ or the critical pressure of $^4$He for $T > 5.2$ K.

Reported theoretical values of $U_b$ for $^4$He on Cu are in the range $-(3.55$ to $6.79)$ meV [59,61,95,96]. More relevant, however, are results for $^4$He adsorption on Cu samples with similar ambient exposures prior to study. For example, isosteric heats of adsorption $Q_{st}$ derived from adsorption isotherms markedly deviate from a constant, increasing from 6.6 meV at $\theta \approx 1$ to about 12 to 15 meV for $\theta \to 0$, which is indicative of inhomogeneous adsorption of $^4$He on Cu [94]. The limit $\theta \to 0$ suggests $U_b \sim -(10$ to $14)$ meV, which is consistent, within experimental uncertainty, with $U_b = -15.2$ meV deduced independently by a time-of-flight diffusion method [97,98].

Inhomogeneous absorption is taken into account by assuming a Gaussian distribution in $U_b$, where the mean $U_{b0}$ and standard deviation $\sigma_b$ are treated as adjustable parameters, giving,

$$\theta = \pi^{-1/2}\sigma_b^{-1} \int_{-\infty}^{U_T} \exp[-0.5(U-U_{b0})^2/\sigma_b^2] \\ \times \ln\{(U/k_B T)/\ln[P/P_{\text{vap}}(T)]\}\, dU + \theta_0, \quad (14)$$

where $U_T = \ln[P/P_{\text{vap}}(T)]k_B T$ and the constant $\theta_0$ is added for general application (Section 3.4). Since the experimental pressure is not given, the function of Eq. (12) is illustrated with $P = 2 \times 10^{-5}$ Torr (twice the experimental base pressure) and setting $\theta_0 = 0$. Parameters obtained from chi-square minimization are $\Delta W = 4.306(2)$ mV, $W_1 = -17.37(8)$ mV, $U_{b0} = -7.60(2)$ meV, and $\sigma_b = 2.84(1)$ meV, yielding a 0.0163 mV rms deviation with respect to the data replica. Fitted functions are shown in Fig. 3, where $W_T - W_{4.2}$ is the dotted curve and $\theta$ is the dashed curve. The chi-square is weakly dependent on the choice of $P$ over the range $10^{-9}$–$10^{-2}$ Torr, owing to its strong correlation with parameter $U_{b0}$. For example, the frequently occurring pressure $P = 8 \times 10^{-9}$ Torr reported by Lockhart et al. [34:Erratum] gives $\Delta W = 4.315(2)$ mV, $W_1 = -26.58(8)$ mV, $U_{b0} = -8.96(2)$ meV, $\sigma_b = 4.69(1)$ meV, and 0.0162 mV rms deviation. At $T = 4.2$ K, the values of $\theta$ are 0.25 and 0.16 at $P = 2 \times 10^{-5}$ and $8 \times 10^{-9}$ Torr, respectively.

### 3.3. Ambient electric field measurement

Experimental methods in the electron-free fall experiments are described in Refs. [30,34], and [41] for measuring the force $F$ on an electron transiting the copper tube by modeling time of flight spectra for electrons emitted from a pulsed source. Analytical functions are based on the Newtonian energy $W_F = F^2 t^2/8m + mh^2/2t^2 + Fh/2$ pertaining to an electron of initial energy $W_F$ with transit time $t$ and traversal distance $h$, and treating $F$ as a fitting parameter [41]. Power-law exponents are used to parameterize the distributions in the electron emission energies and times of delay in traps, together with normalization and background noise parameters [41]. A theoretical maximum in time $t_{max} = (2hm/F)^{1/2}$ at minimum $W_F = Fh$ occurs for $F$ opposing the upward flux of electrons. The ambient force on the electrons was obtained by measuring $F$ with applied electric fields and extrapolating to zero field.

In [30], the negative ambient electric field of about $-5 \times 10^{-10}$ V/m at 4.2 K, observed with vacuum pressure below $4 \times 10^{-10}$ Torr [30], has been argued as correct and unaffected by $^4$He gas scattering [34:Erratum]. Validation that this is the case is the force $F = 1.3 \times 10^{-10}$ eV/m obtained from time of flight distributions in Ref. [31]. The analysis of electron flight times given below makes use of the experimentally verified electron-$^4$He scattering cross section $\sigma_{\text{e-He}}$ at low energy [99], which is smaller by a factor of $2.5 \times 10^{-4}$ than the theoretical value employed in Ref. [41]. Contrary to the negatating conclusions and opinions expressed previously (see, e.g., Refs. [36] and [34:Erratum]), electron-$^4$He scattering proves not to be a valid rationale for retracting the experimental ambient electric fields measured for $T > 4.2$ K.

*3.3.1. Experimental errors*

Scattering of the electrons by ambient $^4$He gas has been presumed in Ref. [34:Erratum] to affect some of the results for $F$ in Ref. [34], although purturbations from gas scattering on time of flight spectra [36] have not previously been explicitly included in analyzing the data. However, errors may be discerned by comparing data from the original experiments (Fig. 2 in Ref. [30]) with claimed gas pressure $P$ well below $10^{-10}$ Torr [34:Erratum] and the heated tube experiments at 4.2 K with $P < 4 \times 10^{-8}$ Torr [34:Erratum] (Fig. 8a in Ref. [100]). The results for the absolute values of $F$ at 4.2 K are combined and plotted against applied electric field $E_A$ in Fig. 4. A solid diagonal line denotes the expression $|F| = |eE_A|$ for zero ambient force. Dotted curves denote the limits of uncertainty for the ambient force, given as $(0.13 \pm 0.51) \times 10^{-11}$ eV/m from statistical analysis of 11 data sets with $|E_A| < 2.5 \times 10^{-10}$ V/m, representing $|F| = |eE_A| \pm 0.09\, m_e g$ [30]. Most of these 11 data points in Fig. 4 are within or near the limits, while several appear to be significant outliers.

The diagonal line in Fig. 4 can essentially be employed as a reference from which deviations from unity of the ratios $r_F \equiv |F/eE_A|$ and $r_{t_{\max}} = r_F^{-1/2}$ provide indications of experimental errors. The 16 data points from Ref. [30] ("WF") exhibit ranges $r_F = 0.24 - 1.42$ and $r_{t_{\max}} = 0.84 - 2.03$; statistics (mean ± standard deviation) are $r_F = 0.98 \pm 0.30$ and $r_{t_{\max}} = 1.07 \pm 0.28$. For the 4 data points from Ref. [100] ("LWF"), the ranges are $r_F = 0.59$ to 1.48 and $r_{t_{\max}} = 0.82$ to 1.30 with statistics $r_F = 0.85 \pm 0.42$ and $r_{t_{\max}} = 1.14 \pm 0.22$. Accuracy of the results for $F$ in experiments with the heated tube at 4.2 K thus appears to be about ± (40 to 50) %, falling outside the experimental error bars of ± (15 to 26) %. Deviations from the diagonal line also suggest systematic underestimations as $|F| \sim 0.6\, |eE_A|$ at the highest applied electric fields $E_A \sim 10^{-7}$ V/m. At elevated tube temperatures of 4.3 to 11 K, ambient electric fields values of $E \sim 10^{-8}$ to $\sim 10^{-6}$ V/m are assigned

± (23 to 53) % errors (Fig. 3 of Ref. [100]), which, in view of the above, are evidently realistic.

Systematic error from gas scattering at a given pressure diminishes with increasing $F$, as the electrons traverse the drift tube with increasing velocities. The induced electric dipole model for electron-$^4$He scattering derived in Ref. [41], evaluated for $P = 4 \times 10^{-8}$ Torr and interaction energy $Fh \approx 10^{-7}$ eV, gives a scattering cross section $\sigma_{\text{e-He}} = 2 \times 10^{-16}$ m$^2$ and a scattering time $\tau_s = 0.04\, t_{\max}$ at 4.2 K. Such analysis portends significant scattering of the electron beam (see, e.g., Refs. [36] and [34:Erratum]) which, in retrospect, was too conservative in view of the directly measured $\sigma_{\text{e-He}} = 5 \times 10^{-20}$ m$^2$ for low energy electrons [99]. At the highest ambient pressure of $4 \times 10^{-8}$ Torr considered possible for a small $^4$He gas leak in Ref. [34:Erratum], ambient fields are therefore reliably obtained for values larger than about $10^{-9}$ V/m in the heated tube experiment [34,100], which is substantially less than the $7.4 \times 10^{-6}$ V/m criterion cited in Ref. [34:Erratum]. The corrected criterion limits observable perturbations by $^4$He gas scattering to data obtained for temperatures near 4.2 K in Refs. [34] and [100].

A significant source of systematic error is the distribution in emission time delays in traps at the electron source, which smear out the cut-offs in the time of flight spectra at $t_{\max}$ [30,41]. Trapping of the electrons prior to their entering the drift tube increases difficulty in data analysis at the larger values of $F$ used in experiments at the elevated drift-tube temperatures [100]. For tube temperatures of 77 and 300 K, results for ambient electric fields were derived from observed changes in intensities of detected electrons with applied electric fields [34,43,100]. Peculiarity in the field dependence, noted for data taken at 300 K in Ref. [42], is considered below for estimating the uncertainty in the reported ambient field.

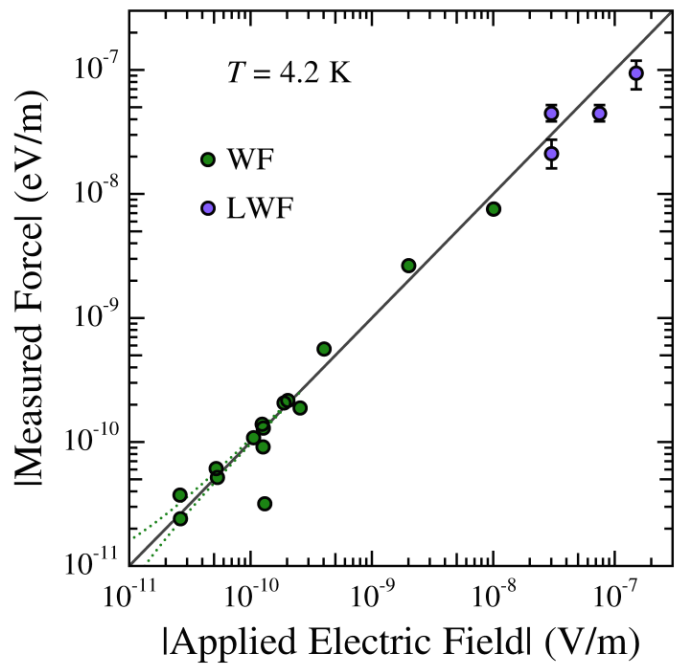

**Fig. 4.** Measured force on free-fall electrons *vs.* applied electric field (absolute values) at temperature $T = 4.2$ K from Refs. [30] (WF) and [100] (LWF). Solid line indicates zero ambient force; dotted curves denote ranges of statistical uncertainty obtained in Ref. [30].

*3.3.2. 300 K data analysis*

An effective ambient electric field of $E = (7.5 \pm 6) \times 10^{-6}$ V/m was measured for the drift tube at 300 K by applying electric fields of magnitudes significantly greater than $E$ [34]. Variation in the relative intensity of detected electrons with applied force $F_A$ is shown in Fig. 5 on linear scales, which transcribe data and trend curves in the original semi-logarithmic plots (see Fig. 7 of Ref. [100]). A dashed line (slope = 0.0116 $\mu V^{-1}$m) illustrates the linear trend pointed out in Ref. [42]. The dotted curve is an estimated trend that is drawn from the experimental description presented in Ref. [34], where the number of detected electrons emitted at energies less than $W$ is determined to be proportional to $W$ itself. The relative intensity can be conjectured from the time of flight expression $W = F^2 t^2/8m + mh^2/2t^2 + Fh/2$ as the ratio $I_r$ of $W$ for $F = F_0 + F_A$ to its value at $F_A = 0$, where $F_0$ is the ambient force ($-eE$) and $F_A$ is the applied force ($-eE_A$). This ratio is approximately represented as,

$$I_r = 1 + F_A h/2W_0 + (F_0 + F_A)^2 h^2/8W_0(2W_0 - F_0 h), \quad (15)$$

where $W_0$ is the value of $W$ at $F = F_0$ and contains implicit dependence on $t$. The dotted curve in Fig. 5 shows Eq. (15) with $F_0 = -7.5 \times 10^{-6}$ eV/m, set to the mean value of $-eE$ in Ref. [34], and fitted to the data with $W_0 = 36 \pm 4$ $\mu$V and normalized to $I_r = 0.97 \pm 0.02$ at $F_A = 0$ (reduced $\chi^2 = 0.88$). Uncertainty in $F_0$ treated as a variable parameter is estimated as $\pm 1.1 \times 10^{-5}$ eV/m, somewhat exceeding the $\pm 6 \times 10^{-6}$ eV/m uncertainty given for $eE$ [34]; $F_0$ acquires large uncertainty owing to the comparatively weak non-linearity in $I_r$ vs. $F_A$ and scatter in the data.

*3.4. Ambient electric field temperature dependence*

Data for the variation in the ambient electric field with temperature obtained in the electron free-fall experiments in Refs. [30,34,101] are plotted in Fig. 6 on linear scales. Circle and triangle symbols distinguish the two methods reported for data analysis [34]. The square symbol is the original result in Ref. [30] with error bars reproduced from later work [34,101]. Ambient fields reported for 77 K [(4 $\pm$ 2) $\times 10^{-6}$ V/m] and 300 K [(7.5 $\pm$ 6) $\times 10^{-6}$ V/m)] were interpreted as measures of the unshielded electric field $E_T$ [34]. The average $E_T$ (= 5.75 $\times 10^{-6}$ V/m) is assumed for the experimentally determined magnitude, notwithstanding a significant 77% uncertainty [34], which is comparable to $M_{Cu}g/e = 6.45 \times 10^{-6}$ V/m and the theoretical estimate given in Ref. [27] for Cu mass $M_{Cu}$.

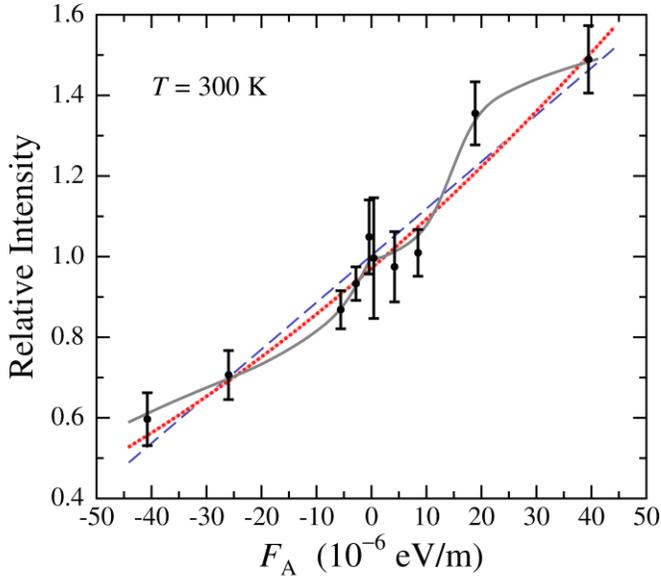

**Fig. 5.** Relative intensity of slow electrons with various applied forces at room temperature, data symbols and solid trend curves from Ref. [100]. Dashed line shows linear trend; dotted curve is fitted model function of Eq. (15).

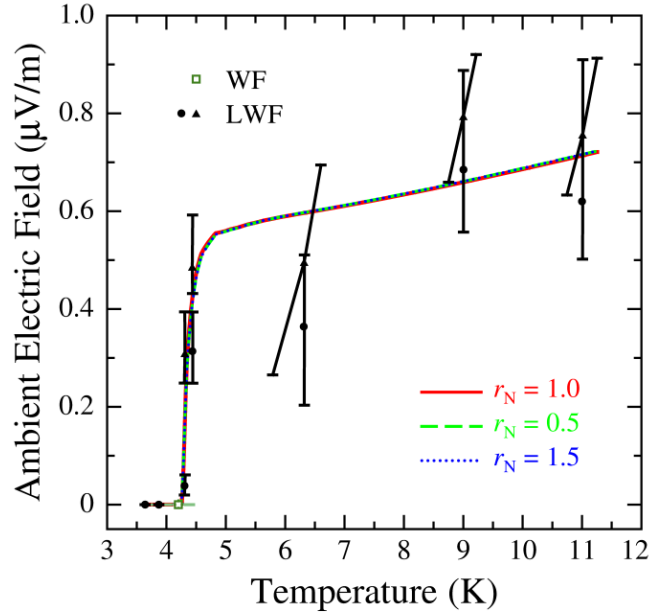

**Fig. 6.** Ambient electric field inside copper tube *vs.* temperature. Filled symbols (LWF) denote data from Refs. [34] and [101]; open square symbol (WF) from Ref. [30]. Solid (red), dashed (green), and dotted (blue) curves are obtained with reduced normal-state resistance $r_N = 1.0$, 0.5, and 1.5, respectively.

The theoretical expression for $E$ in Eq. (9) varies with temperature. Temperature dependence is contained in $\tilde{S}_S$ from Eq. (10), $T_{BKT}$ from Eq. (2), and $T_C = \beta n^{1/2}/\zeta$ from Eq. (1), owing to the temperature dependence in charge density $n$, scaling functionally as

$$n = f_s \theta n_{He}, \qquad (16)$$

where $f_s$ is the fractional number of carriers created per adsorbed $^4$He atom and $n_{He} = 7.9 \times 10^{14}$ cm$^{-2}$ is the monolayer capacity [94]. The fractional monolayer coverage $\theta$ for inhomogeneous $^4$He adsorption is calculated according to Eq. (14), taking values for the binding parameters $U_{b0}$ and $\sigma_b$ from Section 3.2, helium gas pressure $P = 8 \times 10^{-9}$ Torr from Ref. [34:Erratum], and the constant $\theta_0$ as a fitted parameter. Superconductivity is bounded to $T < T_{CT}$ at which $T_C$ equals the temperature as defined in terms of the function,

$$T_{CT} = T_C(T = T_{CT}). \qquad (17)$$

Unknown parameters in the model for ambient electric field $E$ vs. $T$ in Eq. (9) are $r_N$ in Eqs. (4) and (5), $f_s$ in Eq. (16) and $\theta_0$ in Eq. (14). Tube field $E_T = 5.75 \times 10^{-6}$ V/m and gravitational field $E_G = -m_e g/e = -5.58 \times 10^{-11}$ V/m are treated as known (fixed) quantities. Parameter $b = 1$ in Eq. (10) is assumed for consistency with theory [50]. Given the complex nature of the experimental uncertainties described in Ref. [34], the locations of the symbols and the error bars on linear scales are treated as equally weighted data points. Parameters were obtained from non-linear regression of Eq. (9) to the data subject to the constraint that $T_{BKT}$ at $T = 4.2$ K be in the range 4.2 to 4.4 K, as suggested by the experimental temperature transition and which is found to be satisfied for $r_N \in [0.5, 1.5]$. Results for the fitted parameters and $T_{CT}$ for $r_N = 0.5$, 1.0, and 1.5 are listed in Table 1 with uncertainties calculated from chi-square $\chi^2$ analysis [90]; there is a parameter correlation given by $f_s \theta_0 \approx 5.1 \times 10^{-4}$. Parameter $\theta_0$ corresponds to a background charge density $f_s \theta_0 n_{He} = (4 \pm 1) \times 10^{11}$ cm$^{-2}$, such as from a sub-monolayer fraction $\theta_0$ of $^4$He accumulated in cryopumped gas [30]. Dependences of $E$ on $T$ are found to have nearly identical forms for $r_N = 0.5$, 1.0, and 1.5 ($\pm 0.6$% variation in $\chi^2$), as shown by the three nearly overlapping model curves in Fig. 6. Figure 7 gives the corresponding temperature dependence in the fractional $^4$He coverage $\theta$. Table 2 provides values for the temperature-dependent variables $n$, $T_C$, $T_{BKT}$, $T_F$, $\theta$, $\Lambda_L$, $\ell$, $l$, and $k_F \xi_0$ at $T = 4.2$ K for $r_N = 0.5$, 1.0, and 1.5.

Results at the mean $r_N = 1.0$ are $f_s = 0.0598(51)$, $\theta_0 = 0.0086(19)$, and $T_{CT} = 4.826$ K with root-mean-square deviation from the data of $1.52 \times 10^{-7}$ V/m, giving $\chi^2 \approx 1.07$. Data and the corresponding model function (red curve) for the absolute value $|E|$ are plotted in Fig. 8 on the left logarithmic scale, which better illustrates the $10^4$ dynamic range in the data [30,34,101]. In the model curve, solid denotes the superconducting phase ($T < T_{CT}$) and dotted denotes the normal phase ($T \geq T_{CT}$). Variation of $T_C$ and $T_{BKT}$ (right scale) is shown as solid curves for $T < T_{CT}$; the dotted curves are extended to $T > T_{CT}$, illustrating the trends.

**Table 1**

Fitted parameters for the fractional charge per $^4$He atom $f_s$, residual $^4$He coverage $\theta_0$, temperature $T_{CT}$ from Eq. (17), and rms deviation between model function and data for the ambient electric field, for reduced normal-state resistance $r_N = 0.5$, 1.0, and 1.5.

| $r_N$ | $f_s$ | $\theta_0$ | $T_{CT}$ (K) | rms deviation ($\mu$V/m) |
|---|---|---|---|---|
| 0.5 | 0.0511(48) | 0.0101(24) | 4.802 | 0.1533 |
| 1.0 | 0.0598(51) | 0.0086(19) | 4.826 | 0.1522 |
| 1.5 | 0.0691(56) | 0.0074(16) | 4.861 | 0.1513 |

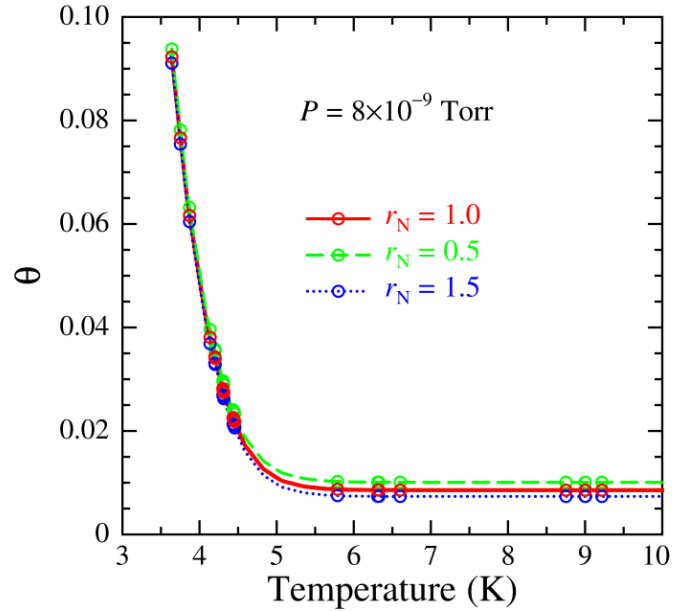

**Fig. 7.** Temperature variation of $^4$He monolayer fraction $\theta$ determined from Eq. (14). Solid (red), dashed (green), and dotted (blue) curves are obtained with reduced normal-state resistance $r_N = 1.0$, 0.5, and 1.5, respectively. Symbols are at the temperatures in the data.

**Table 2**

Parameters of the superconducting layer at $T = 4.2$ K for reduced normal-state resistance $r_N = 0.5$, 1.0, and 1.5: carrier density $n$, transition temperature $T_C$, BKT temperature $T_{BKT}$, Fermi temperature $T_F$, adsorbed $^4$He fraction $\theta$, London penetration depth $\Lambda_L$, carrier spacing $\ell$, mean free path $l$, and product of Fermi wavevector $k_F$ and coherence distance $\xi_0$; parameter $\xi_0$ is constant with temperature.

| $r_N$ | $n$ ($10^{12}$ cm$^{-2}$) | $T_C$ (K) | $T_{BKT}$ (K) | $T_F$ (K) | $\theta$ | $\Lambda_L$ (cm) | $\ell$ (Å) | $l$ (Å) | $k_F \xi_0$ | $\xi_0$ (Å) |
|---|---|---|---|---|---|---|---|---|---|---|
| 0.5 | 1.44 | 7.49 | 4.46 | 58.1 | 0.0357 | 0.270 | 83.2 | 266 | 1.97 | 65.6 |
| 1.0 | 1.61 | 7.92 | 4.40 | 65.0 | 0.0342 | 0.241 | 78.7 | 126 | 2.09 | 65.6 |
| 1.5 | 1.80 | 8.37 | 4.33 | 72.6 | 0.0330 | 0.216 | 74.5 | 79 | 2.21 | 65.6 |

At 4.2 K, the ratio $n/m^* = 2.3 \times 10^{12}$ cm$^{-2}$ $m_e^{-1}$ is ~84% of the value deduced from the effect of $^4$He on the work function, yielding $E_F = 5.6$ meV. Non-universal $\tau_C$ in Eq. (10) is temperature dependent and of magnitude 0.8 at 4.2 K. Reported results for $\tau_C$ give 0.02 to 0.24 for granular films [69,85], 0.21 to 0.47 for amorphous films [85], and 0.3 to greater than 0.9 for gate-charged twisted bilayer graphene [39].

Since $k_F a \approx 0.014$, for (room temperature) lattice constant $a = 4.27$ Å for Cu$_2$O and $k_F = (3.19 \pm 0.02) \times 10^6$ cm$^{-1}$ at $T = 4.2$ K, some of the surface roughness from nanoscale faceting, surface steps, and incommensurate epitaxy tends to be smoothed out. The characteristic length scale $l$ for carrier scattering corresponds to a range of 20 to 60 lattice parameters.

This model obviously presumes that a sub-monolayer of absorbed $^4$He is required for the superconducting shielding, yielding $E = E_G$ at $T \leq 4.2$ K. The abrupt change in temperature dependence for $T > 4.2$ K arises from thermal desorption of $^4$He in combination with fluctuation superconductivity and the transition to the normal state. The form of observed transition in the ambient field is influenced mostly by data at $T \sim 4.2 - 4.4$ K and is for the most part described by the model function curves in Figs. 6 and 8.

## 4. Discussion

The functional dependence of $T_C$ on carrier density, varying as $n^{1/2}$ in Eq. (1), derives from a model of two-dimensional superconductivity mediated by interlayer Coulomb coupling [21,102]. As shown from a pairing ansatz variational model, the zero-temperature gap $\Delta$ also varies as $n^{1/2}$, which is attributed to the constant density of normal electronic states in two-dimensions [1]. These results, obtained from different approaches, evidently constitute equivalent theoretical findings and stand in contrast to three dimensions, where dependence on $n$ appears in an exponent [1].

Equation (1) determines $k_B T_C$ as the product of a strong interlayer Coulomb potential $e^2/\zeta$ and a small interaction factor $\Lambda/\ell$, which evaluate to 7.20 eV and $(9.8 \pm 0.5) \times 10^{-5}$, respectively, for $\zeta = 2.0$ Å and $\ell = 79 \pm 4$ Å (Table 2). These factors derive from the concept of superconductivity propagating via quantum fluctuations within the virtual photon field formed between the two charge reservoirs via a Compton-effective scattering of quasiparticles confined to their respective reservoirs (see, e.g., Ref. [102]), with optimal superconductivity occurring upon charge equilibration between the interacting quasiparticles of the two reservoirs. Asymmetry between (+) and (−) charges is established for Cu/CuO$_2$ by differing dynamics of mobile Cu$_2$O (+) charges and quasi-localized (−) at Cu sites, in addition to the interlayer separation. Thus, pairing occurs between two (+) Fermion charges in the Cu$_2$O layer mediated via exchange of virtual photons with a (−) charge in Cu [21]. While uncertainty exists regarding how the (−) charges are distributed in the Cu, a plausible scenario is a quasi-static distribution of charge reflecting the pattern of adsorbed He atoms. Hence, the (−) charges are quasi-localized in Cu by variable interface bonding potentials, whereas the (+) charges in Cu$_2$O occupy a high mobility interface band. Note also that the mechanism driving the pairing interaction is, by its nature, instantaneous, as in the case treated in Ref. [91] concerning low carrier density superconductors. Unfortunately, there are, at present, no first-principles methods (e.g., based on density functional perturbation theory [103]), since the energies $\sim e^2/\zeta$ of the two virtual photons greatly exceed $E_F$, even with differences in pair momenta small compared to $k_F$.

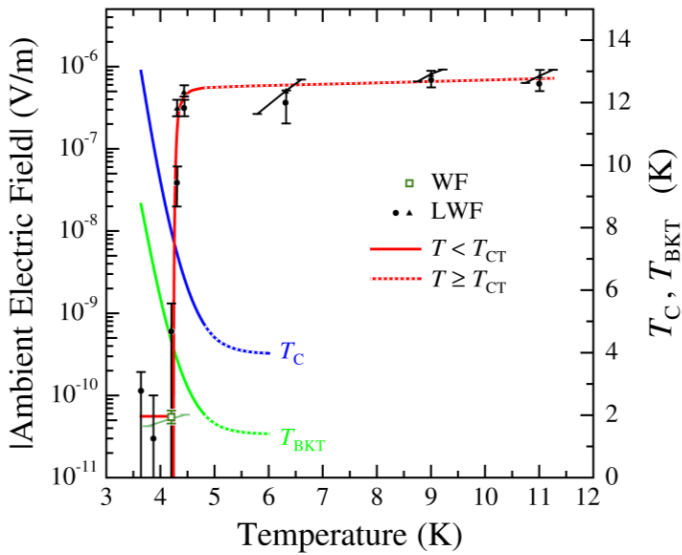

**Fig. 8.** Absolute value of the ambient electric field inside copper tube *vs.* temperature. Filled symbols (LWF) denote data from Refs. [34] and [100]; open square symbol (WF) and vertical error bars from Ref. [30] (slanted error bars from Ref. [34]). The red curve shows the absolute value of ambient field calculated from Eqs. 9 and 10, left scale, with temperature dependences in $T_C$ (blue curve) and $T_{BKT}$ (green curve), right hand scale; superconducting region is denoted by solid curves, normal region by dotted curves.

*4.1. Bosonic effects*

Proximity of the 2D superconductivity to a bosonic pairing phase is determined below, following from parameters determined for Cu/Cu$_2$O with $^4$He adsorption. It may be noted that participation of a fraction of the paired charges in a Bose condensate is sufficient to produce the shielding of ambient electric fields observed in the electron free-fall experiments. For the condensed ideal 2D Bose gas treated in Ref. [12], a theoretical estimate for the shielding factor is $S \approx k_B T/(4\pi e^2 \zeta n N_0)$, where $N_0$ is the ground state occupation number and $\zeta$ is the assumed thickness. An experimental value for the shielding factor at $T = 4.2$ K may be approximated as the ratio of the experimental resolution $0.09 mg/e$ to the unshielded $E_T$, giving $S \sim 8\times10^{-6}$ as an upper limit. This value in the theoretical expression suggests $N_0 \approx 8\times10^3$ as the lower limit. As estimated in Section 2.2, a number $N_V \approx 10^8$ vortices perforate the 2D superconductor, with each core containing about $(n/2)\pi\zeta^2 \approx 4.6$ pairs at 4.2 K. Conjecturing that vortex cores supply a reservoir of non-superconducting bosonic pairs (assuming $\Delta \to 0$ at the vortex core), the observed $S$ is realized with a fraction $\sim10^{-4}$ of core pairs forming a coexisting Bose condensate.

The length $\zeta$ in the Coulomb interaction potential is short relative to the Cooper pair size as well as the inter-particle separation, because the ratio $\zeta/\xi_0 = (\pi/2)^{1/2}\beta k_B m^*/\hbar^2$ acquires the fixed value of 0.0305 in the present case and $k_F\zeta = 0.064 \pm 0.004$ is obtained at $T = 4.2$ K from modeling the ambient electric field data.

Additionally, the superconductive pairing occurs at low carrier density, $k_F\xi_0 = 2.1 \pm 0.1$, which lies near the crossover ($k_F\xi_0 \sim 1$) from superconductivity to Bose-Einstein condensation [1,91,104,105]. Considering that pairing interactions have a relatively short range, useful guidance is provided from calculations of the crossover in 2D that employ short-range or contact interactions and find smooth evolution in the equation of state, pair wavefunction shape, and phase diagram as functions of the interaction parameter $k_F a_{2D}$ ($a_{2D}$ defines the interaction length) [2,9]. Values of $T_{BKT}/T_F$ from Table 2 in combination with the phase diagram for $T_{BKT}/T_F$ vs. $\ln(k_F a_{2D})$ calculated for contact attraction in a 2D Fermi gas [9] lead to the prediction $a_{2D} = 141 \pm 18$ Å, a functional form of $T_{BKT}$ vs. $T_F$ analogous to Eq. (3), and calculated curves for the ambient electric field nearly replicating the results for $r_N \approx 1$ (rms deviation of 0.1518 μV/m with respect to the data).

Equation (3) may be expressed in terms of the Fermi wavevector as,

$$T_{BKT} / T_F = f_C [8\varepsilon_C (1 + 0.75\, r_N k_F \xi_0 /4 )]^{-1}, \qquad (18)$$

and using Eq. (1) to evaluate $f_C = f [(\pi/2)^{1/2} \hbar^2 k_F \zeta\, t_{BKT}/\beta k_B m^*]$

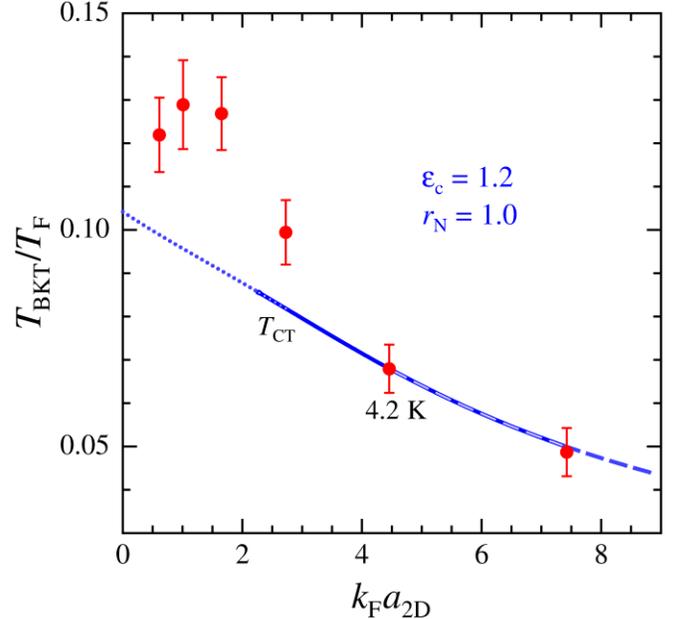

**Fig. 9.** Ratio $T_{BKT} / T_F$ vs. interaction parameter $k_F a_{2D}$ from Eq. (18) for parameters $\varepsilon_C = 1.2$ and $r_N = 1.0$, shown as the solid curve in the temperature range of the ambient field data, dotted overlay for $T > T_{CT}$ and dashed overlay for $T < 4.2$ K; extensions of the curves outside the data range is shown as dotted (left) and dashed (right). Numerical calculations from Ref. [9] are shown as the symbols with error bars.

with the definition $t_{BKT} \equiv T_{BKT}/T_F$. For a fixed $r_N$, as in Section 2.1, this form gives $T_{BKT}/T_F = 1/8\varepsilon_C$ in the dilute limit $k_F \to 0$. The dependence of the ratio for $\varepsilon_C = 1.2$ and $r_N = 1.0$ on $k_F a_{2D}$ is plotted in Fig. 9. The solid underlay curve corresponds to the temperature range 3.64 to 11.25 K in the ambient electric field data, for which $k_F a_{2D} \in$ (2.26, 7.38). The region $T < 4.2$ K ($k_F a_{2D} > 7.38$) is shown with a dashed overlay and the region $T > T_{CT}$ ($k_F a_{2D} <$ 2.26) with a dotted overlay. Two broken curves extending beyond the range of the data are shown as dotted for $k_F a_{2D} \to (\ll 1)$ and as dashed for $k_F a_{2D} \to (\gg 1)$. For comparison, the filled symbols (error bars) show the continuum scaling results from numerical calculations (Fig. 1 in Ref. [9]). The good agreement between the numerical points and the curve in the region corresponding to $T$ up to about 4.2 K underlies the equivalent results in modeling the temperature transition in the ambient electric field discussed in Section 3. It also provides validation of the function of Eq. (1), which predicts that $T_C$ scales with $k_F$. Extension of the range in the function for $T_{BKT}/T_F$ vs. $k_F a_{2D}$ according Eq. (18) shows a linear approach to the upper bound of $1/8\varepsilon_C$ at $k_F a_{2D} \to 0$, whereas the numerical results reach an upper bound of $1/8 \pm 0.008$ for $k_F a_{2D} \sim 1$ and appear to find $\varepsilon_C \to 1 \pm 0.06$ in this regime.

The pair binding energy, defined as $E_b = 4\exp(-2\gamma_E)\hbar^2/m^* a_{2D}^2$ (where $\gamma_E \simeq 0.5771$ is Euler's constant) [2], yields $E_b = 0.7 \pm 0.2$ meV. This result ($E_b/k_B = 8 \pm 2$ K) predicts an appreciable density of bound pairs at $T =$ 4.2 K, where $\ln(k_F a_{2D}) = 1.5 \pm 0.2$, which could allow a minority BEC-like component to coexist and be sustained by the majority superconducting phase for $T < T_{BKT}$. In a mean field model, the binding energy is obtained as $E_b = \Delta_0^2/2E_F$ [1]. Combining Eq. (1), $\Delta_0 = \gamma k_B T_C$, and $E_F = \pi\hbar^2 n/m^*$, one obtains an expression $E_b =$ [3.97(3)] $(\gamma^2\alpha^2/\pi)$ $(m^*/m_e)$ $\hbar^2/2m_e\zeta^2$ which is nominally independent of carrier density, where $\alpha$ is the fine structure constant. Consequently, $a_{2D}$ is predicted to be independent of $n$. Evaluation for $\gamma = 2.5$, $m^* = 0.69\ m_e$, and $\zeta = 2.0$ Å (and $\hbar^2/2m_e\zeta^2 = 0.95$ eV) produces a value smaller than above; $E_b = 0.26$ meV (or $E_b/k_B = 3.0$ K).

*4.2. Weak equivalence principle test*

The original gravitational instrumentation was deemed adaptable for measuring the gravitational force on the positron [31,100] and potentially testing the weak equivalence principle (WEP) of general relativity theory. Note that an argument was posed that Schiff-Barnhill effects render the vertical geometry insensitive for testing the WEP for electrons [106]. Alternative geometries with relativistic or near-relativistic horizontal velocities and non-relativistic vertical velocities have been suggested for improved accuracy in measuring the ratio of inertial and gravitational masses by reducing perturbations by the DMRT and patch fields [107-109]. Near weightless environments are clearly advantageous [106], and a test conducted in Earth's orbit found equivalence to a precision on order $10^{-15}$ [110]. Recent experiments using antihydrogen, which has the advantage of being charge neutral, find downward gravitational acceleration consistent with the WEP [111]; the magnitude for antihydrogen, given as $0.75g$ and uncertainties summing to $0.29g$, is evidently less certain than $(0.98$ to $1.02 \pm 0.09)g$ originally determined for the electron [30,31].

*4.3. Shielding ambient electric fields*

In the proposed model, the superconducting layer on the copper tube yields a shielding factor $S \ll S_N$ at 4.2 K, where the charges induced by adsorbed $^4$He yield $T_{BKT}$ of at least 4.3 K. For $T > 4.2$ K, the value of $T_{BKT}$ falls below 4.2 K, owing to desorption of $^4$He and diminished charge density, such that the onset of superconducting phase fluctuations causes a rapid increase of $S$ towards $S_N$. As $T_C$ drops with increasing temperature, the superconductor undergoes a transition to the normal state at $T_{CT} = T_C = 4.83 \pm 0.03$ K (denoted by the dotted/solid boundary in Fig. 8). Effects of phase fluctuations and $^4$He desorption therefore account for most of abrupt increase in the ambient electric field observed in the heated tube experiment (Figs. 6 and 8) [34,100]. Ambient electric field data for $6\ \mathrm{K} \lesssim T \lesssim 11\ \mathrm{K}$ indicate $S = S_N$ for an electron layer in the normal state, possibly caused by residual $^4$He commingled with other adsorbed ambient gas species. Surface charges as considered in the earlier models [12,43] remain a possibility. Work function data [87,35] confirm the plausibility of superconducting and non-superconducting electronic charges coexisting on the surface of native-oxidized copper and the correlation of induced charge with adsorbed $^4$He.

**5. Conclusion**

Based on previous work on interlayer Coulomb mediation [21,38,102], a model is presented for interfacial superconductivity, arising in a semiconductor-on-metal structure from Coulomb interactions between interface charges in the semiconductor and corresponding screening charges in the metal surface, and considered for the case Cu/Cu$_2$O. The intrinsic superconducting transition temperature $T_C$ for such an interaction scales with the areal density of superconducting charges as $n^{1/2}$ and inversely with the separation $\zeta$ between the two charge layers. Application of this model is offered as an explanation for the shielding of lattice compression and patch electric fields from the copper drift tube observed in the gravitational electron free fall experiments at $T \approx 4.2$ K [30] and the temperature dependence observed in the heated tube experiment [34]. It is also shown that electron-$^4$He scattering is much weaker than previously supposed [34,36] and supports the validity of ambient field data for $T > 4.2$ K as originally reported [34].

Superconducting charges in $Cu_2O$ at the Cu interface and screening charges in the Cu are induced by an inhomogeneous sub-monolayer of adsorbed $^4$He. By modeling temperature-dependent inhomogeneous adsorption of sub-monolayer $^4$He, $n$ and transition temperatures $T_C$ and $T_{BKT}$ become diminishing functions of increasing $T$, owing to $^4$He desorption, and constraining superconductivity to the region $T < 4.83 \pm 0.03$ K. For $\theta \simeq 0.034$ monolayers of $^4$He coverage at $T = 4.2$ K, $n = (1.6 \pm 0.2) \times 10^{12}$ cm$^{-2}$ and theoretically projected $T_C = 7.9 \pm 0.4$ K are calculated for $\zeta = 2.0$ Å, as estimated from interface structure calculations. Finally, it is important to note that $T_C \propto n^{1/2}$ is equivalent to energy gap $\Delta \propto n^{1/2}$ derived for low carrier density superconductors in two dimensions [1].

## CRediT authorship contribution statement

**Dale R. Harshman**: Conceptualization, Data curation, Formal analysis, Investigation, Methodology, Project administration, Resources, Software, Supervision, Validation, Visualization, Writing – original draft, Writing – review & editing. **Anthony T. Fiory**: Conceptualization, Data curation, Formal analysis, Investigation, Methodology, Project administration, Resources, Software, Supervision, Validation, Visualization, Writing – original draft, Writing – review & editing.

## Declaration of Competing Interest

The authors declare that they have no known competing financial interests or personal relationships that could have appeared to influence the work reported in this paper.

## Data availability

Data shown in the figures are available in the cited references.

## Acknowledgements

The authors are grateful for support from the University of Notre Dame, and greatly appreciate communications from F. W. Witteborn and J. M. Lockhart on their readings of an early version of the manuscript.

Corrigendum: D.R. Harshman and A.T. Fiory, Physica C: Supercond. Applic. 632 (2025) 1354600.

The authors regret that misprints are published in the original version of their article. Corrections are presented as follows:

Citation [20,24] in the next paragraph after equation (1) should be [30,34]. Citation [79] in the paragraph preceding section 3.3.1. should be [99]. In the second paragraph after Fig. 5, the exponent −1 at end of the eighth line should be −2, to read cm$^{-2}$. In section 3.3.1, the exponent −11 in the third line before the initial paragraph's end should be −10 to read, $10^{-10}$. Citation [87] in the paragraph preceding section 3.4 should be [34]. Citation [38] ending the second full paragraph after Table 1 should be [39]. Please note that the value 105 assigned to Ref. [105] is out of numerical order in the text. The publication date of Ref. [18] is 2001. The full subtitle of Ref. [51] is "Measurement, Classical Theories and Quantum Theory" and the publication date is 2019. The publication date of Ref. [78] is 1983. The authors apologize for the inconvenience caused.

Concerning Ref. [89], the authors wish to point out that text is truncated in the right margin of the abstract in the printed publication [Bull. Am. Phys. Soc. 23 (1978) 565]. For our readers' convenience, missing text is restored by the authors in the following rendering, as indicated by wavy underline:

---
EN 9  Effect of Helium Gas on the Contact Potential of Copper between 2 and 20K.* J.U. Free,** C.D. Dermer, P.B. Pipes, Dartmouth College.--A non-isothermal Volta technique[1] was used to study the temperature derivative of the contact potential of copper between 2 and 20K in the presence of helium gas. The helium gas pressure was varied between $10^{-6}$ torr and a few mtorr. A sharp change in the contact potential was observed at approximately 4K. The experimental data will be discussed in terms of adsorption models.
* Supported by the Department of Energy
**On leave from Eastern Nazarene College, Quincy, MA
[1] D.H. Darling and P.B. Pipes, Physica 85B, 277 (1977)
---

The direct hyperlinks for certain references in the article are not published online. For readers' convenience, the direct hyperlinks for these articles are presented in Table 1.

Table 1. Reference numbers and the associated hyperlinks to the publications.

| Ref. | Hyperlink to reference article | Ref. | Hyperlink to reference article |
| --- | --- | --- | --- |
| 1 | https://doi.org/10.1103/PhysRevB.41.327 | 63 | https://doi.org/10.1103/PhysRevAccelBeams.22.083101 |
| 4 | https://doi.org/10.1103/PhysRevX.9.031049 | 64 | https://doi.org/10.1063/5.0031568 |
| 7 | https://doi.org/10.1002/adma.202006124 | 67 | https://doi.org/10.1103/PhysRevB.7.3541 |
| 9 | https://doi.org/10.1103/PhysRevLett.129.076403 | 68 | https://doi.org/10.1016/0031-9163(64)90672-9 |
| 12 | https://doi.org/10.1103/PhysRevB.17.1976 | 69 | https://doi.org/10.1103/PhysRevB.27.6691 |
| 13 | http://www.jetp.ras.ru/cgi-bin/dn/e_019_01_0269.pdf | 71 | https://doi.org/10.1103/PhysRevB.69.174505 |
| 15 | https://doi.org/10.1103/PhysRevB.7.1020 | 72 | https://doi.org/10.1103/PhysRevLett.39.1201 |
| 16 | https://doi.org/10.1116/1.1318710 | 73 | https://doi.org/10.1103/PhysRevLett.42.1165 |
| 18 | https://doi.org/10.1103/PhysRevB.64.245112 | 74 | https://doi.org/10.1103/PhysRevLett.56.378 |
| 19 | https://doi.org/10.1380/ejssnt.2004.191 | 75 | https://doi.org/10.1103/PhysRevLett.56.2303 |
| 20 | https://doi.org/10.1103/PhysRevB.71.104510 | 76 | https://doi.org/10.1103/PhysRevB.40.182 |
| 25 | https://doi.org/10.1103/RevModPhys.21.185 | 77 | https://doi.org/10.1103/PhysRevLett.10.486 |
| 26 | https://doi.org/10.1103/PhysRevB.1.4649 | 78 | https://doi.org/10.1103/PhysRevB.27.150 |
| 27 | https://doi.org/10.1103/PhysRev.168.737 | 79 | https://doi.org/10.1103/RevModPhys.36.225 |
| 28 | https://doi.org/10.1103/PhysRev.171.1361 | 81 | https://doi.org/10.1103/PhysRevB.65.144511 |
| 29 | https://doi.org/10.1103/PhysRevB.2.825 | 83 | https://doi.org/10.1103/PhysRevLett.16.308 |
| 30 | https://doi.org/10.1103/PhysRevLett.19.1049 | 84 | https://doi.org/10.1103/PhysRev.164.608 |
| 32 | https://doi.org/10.1103/PhysRevLett.38.1011 | 85 | https://doi.org/10.1103/PhysRevB.28.5075 |
| 33 | https://doi.org/10.1063/1.1686149 | 86 | https://doi.org/10.1103/PhysRev.140.A1197 |
| 34 | https://doi.org/10.1103/PhysRevLett.38.1220 | 87 | https://doi.org/10.1103/PhysRevB.19.631 |
| 36 | https://doi.org/10.1103/RevModPhys.64.237 | 88 | https://doi.org/10.1016/0378-4363(76)90022-X |
| 40 | https://doi.org/10.1103/PhysRev.151.1067 | 93 | https://doi.org/10.2172/5589369 |
| 41 | https://doi.org/10.1063/1.1134860 | 94 | https://doi.org/10.1007/BF00655549 |
| 42 | https://doi.org/10.1103/PhysRevB.17.1934 | 95 | https://doi.org/10.1103/PhysRevB.8.5484 |
| 43 | https://inis.iaea.org/records/jvdvj-n3f86 | 96 | https://doi.org/10.1016/0167-5729(91)90012-M |
| 47 | https://inspirehep.net/files/f55503250f690969aedfda4ceaf9b4f9 | 97 | https://doi.org/10.1103/PhysRevLett.28.346 |
| 48 | https://inspirehep.net/files/0f7b50c47ec26bed99a50ff199960259 | 99 | https://doi.org/10.1007/BF00116695 |
| 50 | https://doi.org/10.1007/BF00116988 | 101 | https://doi.org/10.1088/1361-648X/aa80d0 |
| 55 | https://doi.org/10.1103/RevModPhys.23.203 | 102 | https://doi.org/10.1103/PhysRevLett.125.057001 |
| 57 | https://doi.org/10.1103/PhysRevLett.80.4741 | 103 | https://doi.org/10.1103/PhysRevB.49.6356 |
| 59 | https://doi.org/10.1103/PhysRevB.15.1769 | 104 | https://doi.org/10.1016/j.physc.2023.1354377 |
| 60 | https://doi.org/10.1103/PhysRevB.27.4612 | 110 | https://doi.org/10.1103/PhysRevLett.129.121102 |
| 62 | https://doi.org/10.1103/PhysRevB.3.1215 | | |